\documentclass[manuscript,sigplan,screen,nonacm]{acmart}
\AtBeginDocument{%
  }

\usepackage{tikz}
\usepackage{fancyvrb}
\usetikzlibrary{positioning, arrows.meta, calc}
\usepackage{xcolor}
\usepackage{etoolbox}
\usepackage{longtable}
\usepackage{alltt}
\definecolor{djHdr}{HTML}{CC0000}   % signature / implementation  (hue   0)
\definecolor{djSc}{HTML}{806600}    % constraint 3: the bound      (hue  48)
\definecolor{djSa}{HTML}{007700}    % constraint 1: membership     (hue 120)
\definecolor{djPre}{HTML}{0044CC}   % precondition                 (hue 220)
\definecolor{djSb}{HTML}{AA00AA}    % constraint 2: the quantifier (hue 300)
\newcommand{\dhdr}[1]{\textcolor{djHdr}{#1}}
\newcommand{\dpre}[1]{\textcolor{djPre}{#1}}
\newcommand{\dsa}[1]{\textcolor{djSa}{#1}}
\newcommand{\dsb}[1]{\textcolor{djSb}{#1}}
\newcommand{\dsc}[1]{\textcolor{djSc}{#1}}
\makeatletter
\def\verbatim@font{\normalfont\ttfamily\footnotesize}
\g@addto@macro\UrlBreaks{\UrlOrds}

\renewcommand{\UrlBreaks}{\do\a\do\b\do\c\do\d\do\e\do\f\do\g\do\h\do\i%
  \do\j\do\k\do\l\do\m\do\n\do\o\do\p\do\q\do\r\do\s\do\t\do\u\do\v\do\w%
  \do\x\do\y\do\z\do\A\do\B\do\C\do\D\do\E\do\F\do\G\do\H\do\I\do\J\do\K%
  \do\L\do\M\do\N\do\O\do\P\do\Q\do\R\do\S\do\T\do\U\do\V\do\W\do\X\do\Y%
  \do\Z\do\0\do\1\do\2\do\3\do\4\do\5\do\6\do\7\do\8\do\9\do\/\do\-\do\.}

\usepackage{fancyhdr}
\AtBeginDocument{%
      \addtolength{\footskip}{2.0\baselineskip}%
          \fancyfoot[RO,LE]{Preprint.}%
          }

\begin{document}
\title{Djinnlang: Higher-Level Programming by Unambiguous Specification with an LLM in the Compiler}

\author{Simon Henniger}
\email{shenniger@fas.harvard.edu}
\correspondingauthor
\orcid{0009-0004-0574-2641}
\affiliation{%
  \institution{Harvard University}
  \city{Cambridge}
  \state{Massachusetts}
  \country{USA}
}
\author{Stephen Chong}
\orcid{0000-0002-6734-5383}
\email{chong@seas.harvard.edu}
\affiliation{%
  \institution{Harvard University}
  \city{Cambridge}
  \state{Massachusetts}
  \country{USA}
}
\author{Nada Amin}
\orcid{0000-0002-0830-7248}
\email{namin@seas.harvard.edu}
\affiliation{%
  \institution{Harvard University}
  \city{Cambridge}
  \state{Massachusetts}
  \country{USA}
}

\renewcommand{\shortauthors}{Henniger, Chong and Amin}




\begin{abstract}
  Programmers write formal
  specifications, and LLMs implement them, proving that each
  implementation matches its spec.
  Taken to its extreme, this makes
  specification
  languages the new programming languages.
  We argue that an \emph{unambiguity} constraint is key:
  in addition to proving that its implementation satisfies the specification,
  the LLM must also prove that any other implementation satisfying it
  must produce the same outputs on the same inputs, i.e. that the relation
  formed by the constraints is deterministic.
  This leaves the LLM no leeway on program semantics: as with a conventional
  compiler, the generated code never needs to be read and can be
  regenerated from the spec at any time. Under this constraint and with
  a powerful LLM, the
  difference between a specification language and a programming
  language becomes essentially meaningless, and the LLM essentially becomes a
  part of the compiler toolchain. The arrangement
  doubles as a strong form of AI control: an untrusted model writes the
  code, yet its work is tightly checked by a verifier.

  To demonstrate that our \emph{LLM-in-the-compiler} paradigm is feasible when
  supported by our unambiguity constraint, we present
  \emph{Djinnlang}, a high-level specification language built for this
  future. A Djinnlang program consists only of specifications --- the
  programmer never writes executable code. In place of a traditional
  compiler, a symbolic translator lowers each spec to Dafny stubs and
  proof obligations, and a driver harness orchestrates an LLM that
  fills in implementations and proofs, all checked by the Dafny
  verifier. We evaluate our language and implementation on multiple examples
  and we show that it is self-hosting: an LLM can implement the Djinnlang
  translator from its specification and the reimplementation can verify
  itself.
\end{abstract}

\maketitle
\section{Introduction}\label{introduction}

Here is a plausible future of programming: programmers specify,
LLMs implement and prove their
implementation matches the formal spec.
Taken to its extreme, this scenario has specification languages become
the new programming languages. Similar to high-level programming
languages building on machine instructions, \emph{high-level
specification languages} will abstract upon pure pre/post conditions.

Compared to existing spec languages, we propose to introduce an
\emph{unambiguity} constraint. Under the constraint, a spec will only
yield executable code if it
is unambiguous, i.e. if it fully describes the outputs for any
given set of program inputs. We show that, with the unambiguity
constraint and an LLM implementer, the \emph{difference between a
specification language and a programming language becomes essentially
meaningless}.

We present \emph{Djinnlang}, an attempt to build such a
specification language. With an LLM implementer, it becomes a
general-purpose programming language.
In lieu of a traditional compiler, it comes with a translator
(which symbolically translates the spec into Dafny stubs) and a
driver program (which orchestrates calls to LLMs that write
the actual implementation).
Figure~\ref{fig:max} shows the whole arrangement on a single
example function: five lines of Djinnlang on the left, and on the right the
Dafny obligations our translator emits from them. An LLM will add proofs
and implementation.

\begin{figure*}
\centering
\begin{minipage}[t]{0.31\textwidth}
\scriptsize
\textbf{Djinnlang (written by a human)}
\begin{alltt}
\dhdr{func max(list: seq<int>) -> r: int}
\dpre{requires |list| >= 1:}
  \dsa{r in list}
  \dsb{for x in list:}
    \dsc{x <= r}
\end{alltt}
\end{minipage}\hfill
\begin{minipage}[t]{0.65\textwidth}
\scriptsize
\textbf{Dafny (emitted by \texttt{djinnc})}
\begin{alltt}
\dpre{ghost predicate max_preconditions(list: seq<int>) \{ (|list| >= 1) \}}
ghost predicate max_spec(list: seq<int>, r: int)
\{ max_preconditions(list)
   && (\dsa{(r in list)} && ((\dsb{forall x | x in list ::} \dsc{(x <= r)}) && true)) \}
\textbf{lemma max_is_unambiguous(list: seq<int>, o1: int, o2: int)}
  \textbf{requires max_spec(list, o1)}
  \textbf{requires max_spec(list, o2)}
  \textbf{ensures o1 == o2}
\textbf{\{ /* TODO(LLM): prove. */ \}}
lemma max_exists(list: seq<int>) requires max_preconditions(list)
  ensures exists r: int :: max_spec(list, r)
\{ /* TODO(LLM): prove. */ \}
\dhdr{function max(list: seq<int>): int} requires max_preconditions(list)
\{ /* TODO(LLM): implement. */ \}
lemma max_implementation_is_correct(list: seq<int>)
  requires max_preconditions(list)
  ensures max_spec(list, max(list))
\{ /* TODO(LLM): prove. */ \}
\end{alltt}
\end{minipage}
\caption{A complete Djinnlang function and the Dafny obligations \texttt{djinnc}
  produces from it (predicate names renamed for readability). The left column and the
  two \texttt{ghost predicate}s are trusted: the predicates are a
  mechanical transcription of the constraints.
  Each obligation is pinned, such that a model can only add the bodies
  but never change the definitions. Note
  the unambiguity constraint, in bold, which is described in Section~\ref{unambiguity}.
  Details on the other proof obligations are in Section~\ref{functions}.}
\label{fig:max}
\end{figure*}

Our spec language looks, feels, and behaves like a
higher-than-usual-level programming language. By extension, its
compiler, which includes an LLM and harness, feels (apart from
performance and cost) like a normal compiler toolchain. The verifier
ensures that the LLM's programs satisfy the specification and the
unambiguity constraint ensures that it never has any leeway on semantics. Hence, like
a compiler and unlike in regular vibecoding, there is almost never a
need to read the code it produced, and the code can be regenerated from
the spec at any time. Consequently, we show a new and underexplored
use-case of LLMs in compiler design: when carefully used in tandem with
a verifier, they allow for stronger abstractions.

\newpage

\subsection{Structure and Contributions}
In this paper, we make the following contributions:
\begin{enumerate}
  \item We introduce the unambiguity constraint. We argue that a function should
    be considered unambiguously specified by a set of constraints when the
    relation made up of those constraints 
    is deterministic. We discuss the advantages and limitations
    of this definition (Section~\ref{unambiguity}).
  \item We introduce the
    LLM-in-the-compiler paradigm which puts an LLM into a compiler toolchain
    (Section~\ref{goals}).
  \item Then, we describe Djinnlang,
    a general-purpose language in which programmers write only
    specifications and which uses the unambiguity constraint and LLM-in-the-compiler
    paradigm (Section~\ref{concepts}).
  \item We evaluate the language design qualitatively on specific
    examples which show the language is general-purpose
    (Section~\ref{eval-design}).
  \item We describe the implementation of our Djinnlang compiler pipeline that includes an
  LLM and a verifier, providing a practical example of a compiler that directly
  integrates an LLM (Section~\ref{implementation}).
  \item We evaluate the implementation empirically, measuring
    specification, code, and proof size, proof effort, cost, and the
    performance of the generated programs across two LLM implementers
    (Section~\ref{eval-empirical}).
\end{enumerate}

Finally, we discuss related work (Section~\ref{related-work}), the
limitations of the approach (Section~\ref{limitations}), and
future work (Section~\ref{future-work}), and conclude
(Section~\ref{conclusion}).

% The compilation pipeline figure, \input from overview.md.
% Needs tikz with positioning, arrows.meta, calc (specheader.tex).
%
% Layout: two rows plus a slim LLM row. Programmer and Translator span
% both rows; Harness and Compiler sit in the lower row; the Verifier sits
% in the top row above the Compiler, fed by the spec (top) and the
% implementation (up from the Harness). Nothing leaves the Verifier.
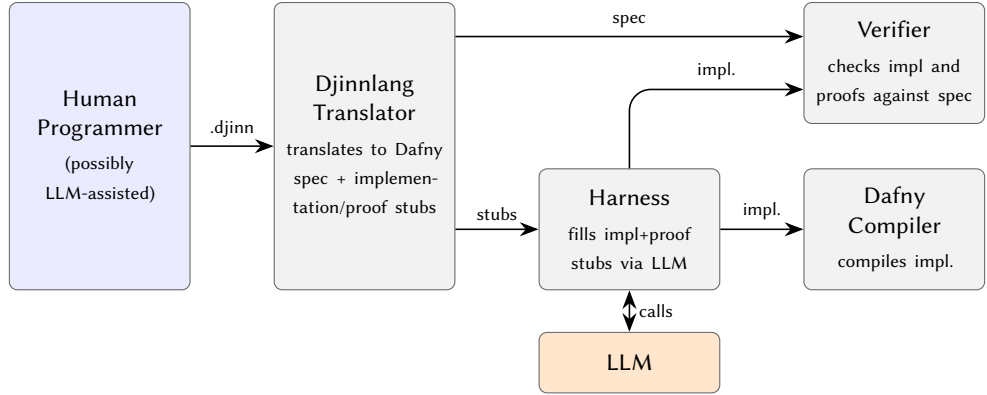
\begin{figure*}[htbp]
\centering
\begin{tikzpicture}[
    font=\small\sffamily,
    box/.style={draw=black!60, rounded corners=3pt, align=center,
                text width=6em, inner sep=4pt},
    tall/.style={box, minimum height=3.8cm},
    short/.style={box, minimum height=1.6cm},
    human/.style={fill=blue!8},
    tool/.style={fill=black!5},
    llm/.style={fill=orange!20},
    lane/.style={-{Stealth[length=2.5mm]}, semithick},
    lbl/.style={font=\scriptsize\sffamily}
  ]
  \node[tall, human] (prog) {Human\\Programmer\\[2pt]
                             {\scriptsize (possibly LLM-assisted)}};
  \node[tall, tool, right=1.1cm of prog] (trans)
                            {Djinnlang\\Translator\\[2pt]
                             {\scriptsize translates to Dafny spec
                              + implementation/proof stubs}};
  \node[short, tool, anchor=west] (harness) at ($(trans.east)+(1.1,-1.1)$)
                            {Harness\\[2pt]
                             {\scriptsize fills impl+proof stubs via LLM}};
  \node[short, tool, right=1.1cm of harness] (dafny)
                            {Dafny\\Compiler\\[2pt]
                             {\scriptsize compiles impl.}};
  \node[short, tool] (verif) at ($(dafny)+(0,2.2)$)
                            {Verifier\\[2pt]
                             {\scriptsize checks impl and proofs against spec}};
  \node[box, llm, minimum height=0.8cm, below=0.55cm of harness] (llm) {LLM};

  % the programmer's source
  \draw[lane] (prog) -- (trans) node[lbl, above, midway] {.djinn};

  % lower row: stubs become the implementation
  \draw[lane] ([yshift=-1.1cm]trans.east) -- (harness.west)
      node[lbl, above, midway] {stubs};
  \draw[lane] (harness) -- (dafny) node[lbl, above, midway] {impl.};

  % top row: spec and implementation meet at the verifier
  \draw[lane] ([yshift=1.45cm]trans.east) -- ([yshift=3.5mm]verif.west)
      node[lbl, above, midway] {spec};
  \draw[lane, rounded corners=8pt]
      (harness.north) |- ([yshift=-3.5mm]verif.west)
      node[lbl, above, pos=0.75] {impl.};

  % the harness drives the LLM
  \draw[{Stealth[length=2.5mm]}-{Stealth[length=2.5mm]}, semithick]
      (harness) -- (llm) node[lbl, right, midway] {calls};
\end{tikzpicture}
\caption{The compilation pipeline. The translator turns the programmer's
  spec into a trusted Dafny spec and implementation stubs; the harness has
  an LLM fill in the stubs; the verifier checks the resulting
  implementation against the Dafny spec; the Dafny compiler receives only
  the implementation.}
\Description{A left-to-right pipeline diagram. The programmer writes a
  Djinnlang specification, which the translator turns into a trusted
  Dafny specification and a set of implementation and proof stubs. The
  harness passes the stubs and the specification to an LLM, which
  fills them in. The Dafny verifier checks the filled-in stubs against
  the trusted specification, returning failures to the harness and the
  LLM in a loop; on success the Dafny compiler turns the
  implementation into an executable.}
\label{fig:pipeline}
\end{figure*}

\section{The Unambiguity Constraint}\label{unambiguity}

Currently, most program properties are not proven by default. We only prove particular
propositions about programs that we care about.

In the LLM era, we propose to turn this principle on its head and instead
unambiguously specify everything about a program by default, except where
explicitly carved out.

This becomes necessary because LLMs are now producing tons of code
with little to no human supervision. But LLM output can fail in two ways: models
misread specifications, hallucinate, and err, and a model may also
subvert its task deliberately --- a possibility safety arguments
increasingly decline to exclude~\cite{greenblatt2024control}. Thus, the LLM must
be tightly constrained. \textit{Even in a context where LLMs write all the code,
they should have no leeway at all on program semantics.} 

\subsection{Deterministic Relations}
Concretely, we propose to consider a function\footnote{Throughout the paper, we use
the term \emph{function} as used in programming, not as used in mathematics. In other
words, a function is a piece of executable code. It is \emph{not a special
case of a relation.} We realize this is initially confusing.} to be
unambiguously specified when the relation
derived only from its specification is deterministic. In other words: to prove a that
a function is unambiguously specified by a set of constraints, we prove
that any two functions that satisfy the set of constraints must return the same outputs
given the same inputs.
Write $\mathit{pre}(\bar{x})$ for the conjunction of a declaration's
preconditions and $q(\bar{x}, o)$ for the relation its constraints
define between inputs $\bar{x}$ and output $o$. The unambiguity
obligation is then
\[
  \forall \bar{x}, o_1, o_2.\;
  \mathit{pre}(\bar{x}) \wedge q(\bar{x}, o_1) \wedge q(\bar{x}, o_2)
  \;\Longrightarrow\; o_1 = o_2 ,
\]
and it comes with an existence obligation,
\[
  \forall \bar{x}.\;
  \mathit{pre}(\bar{x}) \;\Longrightarrow\; \exists o.\; q(\bar{x}, o) .
\]
Once $q$ is a total deterministic relation, an implementation has nothing left to decide
about \emph{what} the program computes.

\subsection{Advantages}
This lets us write programs only as specifications and have LLMs implement them
entirely unsupervised, in the good conscience that our intended program behavior has been entirely
specified and our will (as we wrote it down in the specification) is going to be enforced by the verifier.

A possible objection is that we are merely kicking the can down the road. If programmers
write specifications instead of programs, will these specifications not be equally error-prone
and hard to maintain as current programs?

While we agree that this is by no means the end of all bugs, we think there are a number
of clear advantages to writing unambiguous specifications as opposed to programs.

\paragraph{Contradictions and safety.} Many bugs come down to contradictions and
inconsistencies. One function thinks of something in one way, another
in another, or even one line within a function does it one way,
another in another way. In specs, it is even easier to contradict
yourself. But contradictions are discovered because they make
the spec unsatisfiable, thereby failing at compile time instead of in production.
We argue that, with spec-based programming, the
old adage DRY (don't repeat yourself) will be obsolete: it will be
good practice to repeat yourself, to write out what you have in mind
in as many different ways as possible, because this allows the
verifier to check your work and avoids bugs. Another way of thinking about
this is that, every line in Djinnlang becomes another little piece
that you \textit{know} for certain about your program (because you wrote
it down, an LLM had to prove it for you, and the verifier has confirmed it)---and
more of that is unambiguously better.

\paragraph{Elegance.} Specs need far less operational bookkeeping, and they
often tend to be shorter than the code they replace and closer to how
the programmer already thinks about the problem: \texttt{sort} is a
permutation constraint plus a sortedness constraint---no algorithm
required. As we will show, \texttt{stringtoint} is quickly obtained from
\texttt{inttostring}, as a preimage.

\paragraph{Performance and Universality}. This arrangement makes a human
specify everything that matters to them about code, while leaving all
details up to the LLM. LLMs can do very comprehensive performance
optimizations, and a verifier can check their correctness. We think a
future iteration of this spec language could even be
implementation-language-agnostic.

We show concrete examples of the first two advantages when we evaluate
the language design (Section~\ref{eval-design}); the third is measured
empirically in Section~\ref{eval-empirical}.

\subsection{AI control}
There are concerns about models actively trying to subvert their overseers.
One response to these concerns is AI control, usually by protocols --- trusted
editing, untrusted monitoring --- that lower the probability that undesired
outputs like backdoor slips through safety mechanisms~\cite{greenblatt2024control}.

Our unambiguity constraint could be considered a very thorough type of AI control,
one which not only strategically reduces the model's power, but also provides
a formal guarantee on what it has no power (program outputs) versus some power
(performance) over.

\subsection{Limitations}
Our approach also introduces new dangers. For instance, the specifics
of an LLM-provided implementation
will still determine other factors like performance. This is often
desirable as LLMs are good code optimizers. But of course, it also
introduces additional safety concerns (e.g.~on timing-sensitive
cryptography code or DoS-style attacks with code that is
intentionally slow on some inputs). While highly timing-sensitive cryptography
code (the kind where even performance deviations of a few cycles can lead to successful
side-channel attacks) will likely always remain out of scope for this class of languages
(as they are for many high-level programming languages),
we think there can be some improvements to the current situation.
Future extensions of this
language should include simple syntax for specifying bounds on
baseline performance and scaling behavior and computational complexity.

Another danger is that of bugs in verifiers. If a verifier or another component
of the infrastructure that invokes a verifier does fail, we are at the mercy of the
now fully unsupervised LLM which may be harmful and dangerous in all kinds
of ways. We feel that, in general, our faith in verifiers is justified, but
risks will always remain.

\section{The LLM-in-the-compiler paradigm}
\label{goals}
We propose an \emph{LLM-in-the-compiler} paradigm, which has LLMs become a part
of a compiler toolchain. Unlike existing approaches for LLMs in software engineering,
the outputs of our LLM are tightly verified by other parts of the toolchain. This makes
LLMs feel (apart from performance and cost) no different than traditional program synthesis
algorithms.
This is connected to the unambiguity constraint: only this constraint ensures that
the LLM never has any leeway on semantics. Hence, very much unlike
in regular vibecoding, there is almost never a
need to read the code it produced, and the code is disposable. It can be regenerated from
the spec at any time.

Of course, using an LLM inside a toolchain does not preclude, but rather
complements, other LLM uses on the same code. A programmer writing code
in the spec language may want to \emph{additionally} use an LLM to help
with that task (more in line with standard vibecoding), and all the
benefits of readability and contradiction detection would still apply in this
situation.

Figure~\ref{fig:pipeline} shows the pipeline of our programming language Djinnlang,
which is an example of the LLM-in-the-compiler paradigm.
The pipeline has the shape of an ordinary compiler
toolchain, with one unusual stage in the middle.

The \emph{programmer}, a human, writes a \texttt{.djinn} file containing only
specifications. They may or may not be assisted by an LLM while doing so,
but that is their prerogrative. The specification is the only trusted,
human-generated artifact.

The \emph{translator} is a conventional, symbolic compiler stage. It
parses the source and emits two things. The first is a trusted Dafny
spec: for every \texttt{func}, \texttt{rel}, and \texttt{proc}, the
relation between inputs and outputs that its constraints define, plus
the proof obligations of Section~\ref{unambiguity} --- unambiguity,
existence, and the exported \texttt{ensures} clauses. The second is a
set of stubs: an implementation stub and a proof stub for every
obligation, each with an empty body for the LLM to fill in.

The \emph{harness} helps the LLM do that. It hands the stubs, together
with the spec they refer to, to an LLM and asks it to fill them in with
executable code and proofs. The harness is a driver: it
makes no judgment about the LLM's output and simply passes it on.

The \emph{verifier} is the Dafny verifier. It checks the
filled-in implementation and proofs against the translator's spec. 
When it fails, the
harness returns to the LLM with the verifier's complaints; when it
passes, the \emph{Dafny compiler} receives the implementation and
produces the executable.

\section{Djinnlang}\label{concepts}

After having described our two primary ideas, the unambiguity constraint and the
LLM-in-the-compiler paradigm, we now describe Djinnlang, a proof of concept language that
leverages both.

In Djinnlang, a programmer will only ever write constraints, no code.

In spite of this, we still want Djinnlang to be a general-purpose programming language.
We aim to show that most programs can be somewhat comfortably expressed in this language.
We provide facilities for imperative-style code that interacts with the
outside world and has side effects.

In some sense, Djinnlang is a relational programming language, and it
shares a surface with one: equations rather than assignments,
existential binders, choice, and the ability to run a function
backwards by constraining its output. It differs from other relational
languages, such as Verse~\cite{augustsson2023verse}, in what
it does about ambiguity. A Verse expression may yield many
values and Verse lays choices out in the syntax of the term rather than exploring them
non-deterministically. Djinnlang instead forbids ambiguity (except where it can be resolved in
other parts of the program): a \texttt{func} must be \emph{proved} to relate each
input to at most one output. Verse must find an evaluation strategy that solves the
equations, which its authors leave to future work and which cannot
succeed in general --- by their own account the semantics is
``un-implementable'', since no rewrite system will find the solution
to $\exists x.\ x^2 - x - 6 = 0$. That is exactly the burden we hand
to an LLM, which is unpredictable but nearly universal, and check with
a verifier.

To focus on the novel aspects of this language, we piggyback on Dafny for
everything else. Djinnlang defines no type system and no standard library
of its own: any Dafny type may be used in signatures, any function or
predicate from the ambient Dafny context may be called from constraint
expressions, and Dafny's built-in vocabulary (cardinality
\texttt{\textbar{}s\textbar{}}, indexing \texttt{s{[}i{]}}, slicing
\texttt{s{[}a..b{]}}, membership \texttt{x in s}, \ldots) is used
directly in expressions.
We are hoping to create variants that build on Verus instead of Dafny,
as well as on emerging Lean-based code verification systems.

\iffalse
]}}]]]]]]]]}}}}}}}
\fi

There are three top-level constructs: functions, procedures, and
relations. Functions (\texttt{func}) become Dafny functions. They are
guaranteed to be pure (no side effects) and to be unambiguous
(deterministic). Procedures and relations each drop one of these
guarantees:
Procedures are unambiguous, but can have side effects. Relations, on the
other hand, are pure, but can be nondeterministic; to Dafny, they are
predicates and carry no executable code.

Within a function, relation, or procedure, each line is a constraint.
Each states a fact relating the inputs to the output, and the body means
the conjunction of its lines:

\begin{verbatim}
func max(list: seq<int>) -> r: int
requires |list| >= 1:
  r in list                 # r is one of the elements ...
  for x in list:
    x <= r                  # ... and none of them is larger
\end{verbatim}

Nothing here says how to find the maximum---no accumulator is carried
and nothing is compared in any particular order. The two lines only
pin down which output is acceptable, and between them they admit
exactly one; the LLM supplies an implementation and proves it meets
them.

Headers may carry \texttt{requires} and \texttt{ensures} annotations:

\begin{verbatim}
func naturalnumbertostring(i: int) -> s: string
requires i >= 0    # callers prove this at every call
ensures |s| >= 1:  # exported to callers; must follow
  ...              # from the body
\end{verbatim}

Every \texttt{func} and \texttt{proc}
implicitly includes an existence proof (for every valid input,
an output exists) and a terminating implementation; a contradictory spec
is \emph{rejected}. A \texttt{rel}, on the other hand, may
be partial (\texttt{total} opts in to realizability), and a
\texttt{proc} may be marked \texttt{noreturn} when it is meant to run
forever (an event loop) --- everything else provably halts.

\subsection{Relations in Imperative Clothing}\label{clothing}

Note that \texttt{=} is equality, not assignment. \texttt{s = f(x)}
and \texttt{f(x) = s} mean the same thing, so constraints can be
written ``backwards'' --- here parsing is specified in terms of
printing:

\begin{verbatim}
func stringtoint(s: string) -> r: int:
  if s in range(inttostring):
    inttostring(r) = s      # r is whatever prints as s
  else:
    r = 0                   # error handling
\end{verbatim}

The same is true for other constructs.
Djinnlang's syntax is borrowed from structured imperative programming
--- indentation-delimited blocks, \texttt{if}/\texttt{else},
\texttt{for}, one statement to a line, infix \texttt{=}. But none of it executes.
Instead, a body is the conjunction of its lines (which can thus be in any order, with
the exception of \texttt{do} statements in \texttt{proc}, see below).
\texttt{if C: B else: E} is $C \Rightarrow B$ with
$\lnot C \Rightarrow E$, and \texttt{for x in xs: B} is
$\forall x \in xs.\;B$.

Oftentimes, this is very intuitive and makes the specifications look like
imperative programs. For instance, here is the rulebook of Sudoku:

\begin{verbatim}

rel completion(g: seq<int>) -> c: seq<int>
requires |g| = 81:
  |c| = 81
  for i in 0..81:
    1 <= c[i] <= 9
    g[i] != 0 => c[i] = g[i]
  for i in 0..81:
    for j in 0..81:
      i != j && sameunit(i, j) => c[i] != c[j]
\end{verbatim}

Read imperatively, this is the validity checker one would write by
hand: walk the cells, check each digit is in range and agrees with any
given, then walk the pairs and check that no two cells sharing a unit
collide. Read logically, the same text defines the set of solved
grids, and a solver is obtained by saying its answer belongs to that
set. The checker is not written twice, and the nested loop that an
imperative reader sees --- 6561 iterations --- is a quantifier that
costs nothing and constrains no algorithm; the implementation may
propagate constraints and never enumerate a pair.

The disguise keeps specifications skimmable by readers not fluent in
logic, which matters when the specification is the only artifact
anyone reads, and it suits an implementer who is unfamiliar with logic
or relational programming.

However, it can sometimes mislead. We acknowledge that the syntax can appear
confusing and alien at first, similar to how an imperative programmer might
initially struggle to read a functional program.
For instance, lines may look sequential
but hold simultaneously, so a reader tracing data flow looks for an order that
is not there. A \texttt{for} that a programmer reads as accumulating
cannot accumulate, since there is no state to carry; inductive
definitions must be recursive. An \texttt{if} without an \texttt{else}
does not mean ``do nothing otherwise'' but leaves the output
unconstrained, which would often be rejected as ambiguity.

\subsection{Functions}\label{functions}

We will now discuss the three types of top-level blocks in depth, beginning
with functions.

A \texttt{func} is translated to a Dafny function. It must be specified
unambiguously and must not have side effects.

For a \texttt{func f(xs) -> o: T} the translator emits, into the
trusted file, a predicate \texttt{f\_pre(xs)} conjoining the
\texttt{requires} clauses and a predicate \texttt{f\_spec(xs, o)}
holding exactly when the body's constraints relate the inputs to the
output. Into the solution file it emits the following obligations, in
this order, each with an empty body:

\begin{enumerate}
\item \emph{unambiguity} --- for fixed inputs any two outputs
  satisfying \texttt{f\_spec} are equal. This is unambiguousness: the
  relation is a function.
\item \emph{postconditions} --- one lemma per \texttt{ensures} clause,
  deriving it from \texttt{f\_spec}. A postcondition that does not
  follow from the body is rejected rather than assumed.
\item \emph{existence} --- some output satisfies \texttt{f\_spec} for
  every input admitted by \texttt{f\_pre}. A contradictory
  specification dies here.
\item \emph{implementation} --- a compiled, terminating Dafny function
  with the same signature. This is the only stub that is code rather
  than proof.
\item \emph{implementation correctness} --- the implementation's
  output satisfies \texttt{f\_spec}, unconditionally on the
  precondition domain.
\item \emph{call sites} --- where the body calls something with a
  \texttt{requires}, its precondition holds under everything known at
  that point. Omitted when the body incurs none.
\end{enumerate}

Find an example function with all
proof and implementation obligations in Figure~\ref{fig:max}.

\subsection{Relations}\label{relations}

A \texttt{rel} is only a set of constraints. Its specification could be ambiguous and
thus the function nondeterministic. It is not directly implemented and
never executable. It provides loose vocabulary for other specs to
refine. A \texttt{rel} can only be used inside a \texttt{func} and
\texttt{proc} and it is the user's job to resolve any ambiguity.

Calls to relations cannot have side effects.

\begin{verbatim}
# one input, many outputs: fine
rel perm(list: seq<int>) -> t: seq<int>:
  |t| = |list|
  for x in list:
    multiset(t)[x] = multiset(list)[x]
\end{verbatim}

With \texttt{perm}, the \texttt{sort} above can be written as
\texttt{perm(list) = r} plus the ascending constraint. With the latter,
\texttt{sort} sort will be unambiguously specified even when \texttt{perm} is not.

\paragraph{Partial and total relations.}
A \texttt{func} is total by construction: it must relate every input
admitted by its preconditions to some output, and that is one of the
obligations it owes. A \texttt{rel} is not. Left unannotated, a
relation is \emph{partial}: it may relate an input to no output at
all, and the language asks for no proof that it does. Annotating it
\texttt{total} adds exactly the existence obligation a \texttt{func}
carries, $\mathit{pre}(\bar{x}) \Rightarrow \exists o.\;
q(\bar{x}, o)$.

Similar to unambiguousness, relations could be partial, but this partiality
needs to be resolved in whichever functions they are used.
Marking a relation \texttt{total} resolves this problem, which will often
make proofs at the call-site easier.

We chose partiality as the default because a relation's domain is
frequently the point of writing one. A relation with no output on some
inputs is a predicate in disguise, and the language exposes it as one:
\texttt{x in domain(q)} holds exactly when \texttt{q} accepts
\texttt{x} and relates it to at least one output, so a specification
can branch on solvability without ever exhibiting a solution. The
Sudoku specification of Section~\ref{eval-programs} is written this
way --- a puzzle has a solution precisely when it lies in the domain
of the completion relation, and the answer must be a completion
exactly when that test succeeds --- and the same device gives
substring search as membership in the domain of a match-position
relation. For a \texttt{func}, or for a relation marked \texttt{total},
the test degenerates to the precondition, since the output is
guaranteed.

\subsection{Procedures}\label{procedures}

A \texttt{proc} is where the language meets the outside world, and it
was the hardest part of the design to settle. The difficulty is that a
specification is a set of simultaneous constraints while effects
happen one after another: nothing in a conjunction says which of two
effects comes first.

We have settled on an algebraic effects system that expresses imperative
code as a step function.
As an illustrative example, printing two words
becomes a function from the last completed step to the next one,

\begin{verbatim}
func hello_step(last: StepResult) -> s: Step:
  if last = Initial:       s = PrintHello
  else if last = Hello:    s = PrintWorld
  else if last = World:    s = Return
\end{verbatim}

together with a driver written directly in Dafny:

\begin{verbatim}
method hello_world() {
  var last := Initial;
  while true {
    match hello_step(last) {
      case PrintHello => print("Hello"); last := Hello;
      case PrintWorld => print("World"); last := World;
      case Return     => return;
    }
  }
}
\end{verbatim}

\texttt{proc}s are syntactic sugar for this above construct. Whenever a
specification contains a \texttt{proc}, the translator automatically creates
the step function (which the LLM must implement and which must be unambiguously
specified like any other function) and a Dafny driver function. Because the
Dafny driver function is mechanically generated by the translator and never touched
by the LLM, we can trust it.

Within a proc, each step becomes a \texttt{do} instruction.

\begin{verbatim}
proc hello_world():
  do print("Hello")
  do print("World")
  # implicit return; `return` may also be written
\end{verbatim}

Not only can we call functions, but we can also use and change local and
global variables, e.g. by writing \texttt{do myvariable := 42}.

As seen in our example above, \texttt{do} calls are chained together by
the order in which they appear, making \texttt{do} the big exception in
a language in which order generally does not matter.
This trivially works across \texttt{if} statements.  An
\texttt{if} condition is conjoined onto the
guards of the events inside, whose predecessors remain whatever
precedes the \texttt{if}.

For loops, a loop's last event is chained either to the
first event of the next iteration or to whatever follows the loop,
according to the loop condition. Because all clauses sharing a
predecessor carry mutually exclusive guards, the machine is
deterministic by construction and the translator can discharge that
proof itself.

On top for \texttt{for}, procedures also allow for \texttt{while} loops,
where the loop condition is always a step and can thus depend on outside
factors that change during the loop.

\texttt{for} loops can contain \texttt{do} statements, but then they end
up imposing an execution order, which must be syntactically marked by using
\texttt{ordered for}. Not every valid \texttt{for} condition is also a valid
\texttt{ordered for} condition: if a set that is iterated over has no inherent
order, \texttt{ordered for} cannot be used as it would be ambiguous.
\texttt{ordered for} is sugar for a \texttt{while} with
an assigned index: bounds are evaluated once on entry, and the index
is readable in the body, not reassignable, and out of scope afterwards.

\begin{verbatim}
proc printstring(s: seq<char>) -> r: int:
  ordered for i in 0..|s|:
    do putchar(s[i])       # one event per character, in order of i
  r = |s|
\end{verbatim}

The pure constraints in a \texttt{proc} behave as they would in a
\texttt{func}: they hold throughout the run rather than at the point
where they are written. This can initially be confusing. In
\texttt{do x := "10"}, then \texttt{a = stringtoint(x)}, then
\texttt{do x := "hello"}, the middle line is not a step but a claim
about every value \texttt{x} ever takes, so the specification is
contradictory and is rejected. To avoid this, constraints can be scoped
to only hold after certain steps using a
\texttt{then:} block, whose constraints bind only for its extent and
whose relational variables are scoped to it; it must immediately
follow a \texttt{do}, a \texttt{while}, or an \texttt{ordered for}.

Beyond what a \texttt{func} gives, a \texttt{proc} guarantees that
events occur in the specified order, because every event's guard names
its predecessor, and that state changes only where directed, because
the driver hands the state tuple back verbatim and never inspects it.
Contract proofs use an implementer-chosen \emph{machine invariant}
over that state. By default, each \texttt{proc} is also proven to terminate,
though this can be disabled using the \texttt{noreturn} attribute.
We aim to implement higher-level specifications about the event trace
(Section~\ref{future-work}).

\section{Evaluating the Language Design}\label{eval-design}

We evaluate Djinnlang against the goals set out in the introduction.
We aim to show that it is \emph{elegant} by presenting examples in
which the spec is shorter and closer to the programmer's intent than
the code it replaces --- sorting (Section~\ref{eval-sort}), and
\texttt{stringtoint} obtained as a preimage of \texttt{inttostring}
(Section~\ref{eval-preimage}). Between the two we show that deliberate
redundancy lets the verifier catch realistic mistakes in a
specification of integer printing (Section~\ref{eval-redundancy}).
The \emph{general-purpose}
design principle is borne out by the compiler itself, which is written
in Djinnlang (Section~\ref{eval-selfhosting}). For the remaining
advantage, \emph{performance}, we refer to Section~\ref{eval-empirical}.

\subsection{Sorting a list}\label{eval-sort}

The canonical example for any specification language is sorting.
We first define a \texttt{rel} which maps a list to each of its permutations
and then use it:

\begin{verbatim}
rel perm(list: seq<int>) -> t: seq<int>
total:
  |t| = |list|
  for x in list:
    multiset(t)[x] = multiset(list)[x]

func sort(list: seq<int>) -> r: seq<int>:
  perm(list) = r          # (1) r is SOME permutation...
  # You can conceptualize this as sets and
  # read this first line as:
  # r in set_of_all_permutations(list)
  for i in 0..|r|-1:
    r[i] <= r[i+1]        # (2) ... namely a sorted one,
                          # resolving perm's ambiguity
\end{verbatim}

Note that, in order to show \texttt{sort} is deterministic, the LLM
implementer must show that \texttt{\textless{}=} is a total order on
\texttt{int}. If we attempt this on a different type for which this is
not the case, the verifier will flag the issue.

\subsection{int to string}\label{eval-redundancy}

You can implement \texttt{inttostring} inductively or iteratively. The inductive
implementation is elegant, but we still show an iterative implementation (although
both are logically equivalent and thus will not lead to programs of different speeds),
because it helps us demontrate how deliberate redundancy
lets the verifier catch mistakes. We implement it as two functions, first
\texttt{nattostring} and then a small wrapper that handles the sign. The
first function:

\begin{verbatim}
func nattostring(i: int) -> s: string
requires i >= 0:
  |s| = (if i = 0 then 1 else log(10, i) + 1)
  for k in 0..|s|:
    s[|s|-1-k] = ('0' as int + i/pow(10,k) % 10) as char
    s[|s|-1-k] >= '0' # DO REPEAT YOURSELF. Challenge
    s[|s|-1-k] <= '9' # your beliefs about your code by
                      # writing them out.
  i = 5 => s = "5"        # read: if i = 5 then s = "5"
  i = 128 => s = "128"
\end{verbatim}

While this is superficially similar to a direct implementation (apart from
missing housekeeping like explicitly resizing the target string), there
are some benefits: first, you'll note that the last four lines are not
actually necessary. They are an example of ``\emph{do} repeat yourself''
and seamlessly introduce constraints that the verifier checks. Second
and more importantly, note that this is a bit hairy and many small
mistakes could occur. Some examples:

\begin{itemize}
\item
  For instance, let's suppose that we write \texttt{\% 100} instead of
  \texttt{\% 10}. This would trivially be caught by the verifier in the
  \texttt{s{[}\textbar{}s\textbar{}-1-k{]} \textless{}= \textquotesingle{}9\textquotesingle{}}
  constraint.
\item
  Suppose the index had been written
  \texttt{s{[}\textbar{}s\textbar{}-k{]}} rather than
  \texttt{s{[}\textbar{}s\textbar{}-1-k{]}}. At \texttt{k=0} that
  constrains \texttt{s{[}\textbar{}s\textbar{}{]}}, which is out of
  bounds and caught by the verifier; \texttt{s{[}0{]}} would also be
  left unconstrained, so determinism would fail too.
\item
  The off-by-one in the other direction,
  \texttt{s{[}\textbar{}s\textbar{}-2-k{]}}, is caught symmetrically:
  for \texttt{k=\textbar{}s\textbar{}-1}, we would constrain
  \texttt{s{[}-1{]}}, again out of bounds, and
  \texttt{s{[}\textbar{}s\textbar{}-1{]}} would additionally be left
  unconstrained.
\end{itemize}

The second function, \texttt{inttostring} is easy---it introduces a
``-'' if the number is negative. Note that the verifier must prove that
it will never call \texttt{nattostring} on a negative number as that is
a precondition.

\begin{verbatim}
func inttostring(i: int) -> s: string:
  s = (if i >= 0 then nattostring(i) else "-" + nattostring(-i))
\end{verbatim}

\subsection{string to int, as a preimage}\label{eval-preimage}

By using the relational properties of \texttt{inttostring}, we can
directly define \texttt{stringtoint}:

\begin{verbatim}
func stringtoint(s: string) -> r: Result<int>:
  if s in range(inttostring):
    r.ok()
    inttostring(r.value) = s    # r.value is a preimage of s
  else:
    r = Failure
\end{verbatim}

Note that this function is very strict! It only accepts integers exactly
as \texttt{inttostring} would generate them. If we wanted to accept
whitespace, we could use the same pattern, but use the preimage of a
relation which maps every integer to multiple possible string
representations.

Here is a start (only implementing leading
whitespace):

\begin{verbatim}
rel nattomanystrings(i: int) -> s: string
requires i >= 0
total:
  # w is the number of leading whitespace bytes
  0 <= w < |s|
  for k in 0..w:
    s[k] = ' ' || s[k] = '\n' || s[k] = '\t'
  
  |s| = w + (if i = 0 then 1 else log(10, i) + 1)
  for k in 0..|s|-w:
    s[|s|-1-k] = ('0' as int + i/pow(10,k) % 10) as char
    s[|s|-1-k] >= '0'
    s[|s|-1-k] <= '9'
\end{verbatim}

Now we can define:

\begin{verbatim}
func stringtonat(s: string) -> r: Result<int>:
  if s in range(nattomanystrings):
    r.ok()
    nattomanystrings(r.value) = s
  else:
    r = Failure
\end{verbatim}

and \texttt{stringtonat("    12")} will work. Suppose we now
retire \texttt{nattostring}, keeping \texttt{nattomanystrings} as
the single source of truth for digit rendering, and rewrite
\texttt{inttostring} on top of it. Since \texttt{nattomanystrings}
is a relation, \texttt{inttostring} must either become one itself
or resolve the ambiguity. Let us resolve it:

\begin{verbatim}
func inttostring(i: int) -> s: string:
  if i >= 0:
    s = nattomanystrings(i)
    s[0] >= '0' # resolve ambiguity! no leading 
    s[0] <= '9' # whitespace in the string
  else:
    s = "-" + nattomanystrings(-i)
    s[1] >= '0' && s[1] <= '9'
\end{verbatim}

% add 2-3 more nice examples from our other programs here

\section{Implementation}\label{implementation}

Figure~\ref{fig:pipeline} gives the vague shape of the toolchain. In this
section, we describe individual components more concretely. We have primarily
implemented what we call the translator, \texttt{djinnc}, and the rest is
off the shelf: Dafny verifies, Dafny's C\# backend compiles, and the
agent is an unmodified coding assistant.

\subsection{Translator}
Our translator has two modes. The first parses a specification and
writes two files: \texttt{spec.dfy}, the trusted lowering, which holds
a precondition predicate and a relation for every declaration and, for
every \texttt{proc}, the datatypes and step relation of its compiled
machine; and \texttt{solution.dfy}, the obligations, each an
empty-bodied lemma or function stub. It then runs
Dafny on \texttt{spec.dfy} alone, which is how a Djinnlang program is
type-checked --- the language has no checker of its own. Since
\texttt{solution.dfy} holds work that is costly to reproduce, it is
written only when absent, or when forced.

This output is meant to be given to an LLM.
Once the LLM is done, we call the second mode.
\texttt{verify} first
\emph{retranslates} the specification, so that a stale or tampered
\texttt{spec.dfy} can never be what the solution is checked against.
It then scans the solution for escape hatches --- the substrings
\texttt{assume}, \texttt{\{:axiom\}}, \texttt{\{:extern\}}, and
\texttt{\{:verify} --- and checks that every regenerated obligation
appears in the solution \emph{verbatim}: the comment, the
signature, and the contract, everything except the body between the
braces. If all of that passes, it invokes Dafny, with a 120-second per-obligation
limit.

\subsection{Harness}
Our approach is harness-agnostic. For this experiment,
we use standard coding agents such as
Claude Code and Codex. We point them at the
output directory which includes: the specification and its
ambient Dafny, the stub file, and a written guide explaining what each
obligation kind means, how the verifier judges, and which Dafny proof tactics
tend to work on these goals. The
agent's loop is the obvious one --- edit, run \texttt{djinnc verify},
read the errors, repeat --- and it terminates when its implementation passes.

Using an off-the-shelf agent makes this project economically feasible.
Since we use Claude Code and Codex and not API models directly, we can
put our LLM usage on the flatrate subscriptions offered by major AI companies and
do not pay per call.

\subsection{Support tooling}
We were pleasantly surprised to see that our agents have developed some support
tooling for themselves that is not part of the trusted base.
A whole program at once is more than an agent's context comfortably
holds, so they partition the declarations into disjoint groups and give
each group to its own agent working on its own copy of the solution.
A merge tool then reassembles one solution from the copies.

\subsection{Hillclimbing performance}
Semantics is settled by the specification, but speed is not, so
performance is a second, separate pass. Once we have a passing solution,
we can ask an agent in a second prompt to make it faster
under the same obligations. In this round, we only accept the agent's
changes if the proofs still pass verification and the solution is
actually faster across the workloads.

\subsection{Bootstrapping}\label{eval-selfhosting}
\texttt{djinnc} is itself written in Djinnlang (with some pieces of Dafny and C\#
for the environment) and bootstrapped.

To bootstrap, we start out with the specification in Djinnlang. To get
obligations out of it at all, we had an LLM write a throwaway seed translator in
Python (whose output we checked manually). A second LLM filled the
implementation and proof bodies the seed emitted, and the binary built
from those bodies then regenerates its own
obligations byte-for-byte and re-runs the entire proof through its own
\texttt{verify} command --- retiring the seed as a translator and as a
verifier respectively.

\texttt{djinnc} is 4,088 lines of Djinnlang in 487 declarations, from which
it emits 2,106 proof obligations; discharging them took
4,715 lines of implementation and 31,823 lines of proof, together with
1,549 helper declarations. The implementation of procedures makes up
a fair amount of the specification, as do error messages. We
did not optimize the compiler specification for brevity, but instead
have repeated ourselves many times.
Producing the implementation cost GPT~6 Astra 1.4 billion tokens
(an estimated \$1,950 if used via the API, but we ran this entirely through
the flatrate subscription) over multiple days.
While we were still drafting the specification, the compiler found four
distinct bugs in it --- in each case an obligation that could not be
discharged because the draft was genuinely wrong, not merely hard.
Once the specification was corrected, synthesis of the whole compiler
succeeded on the first attempt.
In the following section, we show more examples and
analyze performance more deeply.

\section{Empirical Evaluation}\label{eval-empirical}

Given our implementation and language definition, we now ask: are LLMs
actually good enough to do what we ask of them, and how long does it take
them? We fixed three
specifications and had two different implementers compile them: Claude
Fable 5.1 driven by Claude Code, and GPT~6 Astra driven by Codex.
Neither was allowed to change a specification. Each was asked to
produce a verified implementation, snapshot it, and then perform a
single round of optimization, so that every specification yields four
artifacts, two for each implementer.

All of it worked the first time. Both
implementers discharged all 256 obligations of all three
specifications, and the compiler's 2,106 obligations
(Section~\ref{eval-selfhosting}) came out of a single run as well.
Of course, within a run the agent naturally iterates against the verifier, as a
human proof engineer would.

\subsection{Benchmark programs}\label{eval-programs}

The three programs were chosen so that they can be somewhat elegantly specified,
have CPU-bound workflows, and
the same specification admits both an obvious slow implementation and
a much faster one.

\emph{Sudoku} (70 lines of Djinnlang, 61 obligations) specifies the
lexicographically least completion of a partially filled grid, or a proof that none
exists. Completions are a relation, since a puzzle may have none or
many; minimality is pinned down without quantifying over grids by
requiring that for every empty cell and every smaller digit, the
puzzle that keeps the answer's prefix and places that digit has no
completion at all.

\emph{LZ77} (81 lines, 63 obligations) specifies compression with
optimal parsing. A \texttt{decode} relation fixes the format, and the
minimum token count is given as a recurrence over the finite menu of
tokens usable at a position, which states ``shortest among all
encodings'' without naming an encoding. Note that the compression ratio is
therefore a property of the specification: implementations may differ
only in speed.

\emph{Imp VM} (166 lines, 132 obligations) specifies the small-step
semantics of a small imperative language with unbounded integers,
heap-allocated arrays, loops, and recursive procedures, run under a
fuel bound so that the interpreter is total. An implementation may
walk the syntax tree or compile to a register machine, provided it
stops at exactly the same step count when the fuel runs out.

The specifications were themselves LLM-generated, but we checked them
carefully.

Correctness was checked independently of the proofs by an LLM-generated
black-box test suite --- 14, 30, and 65 cases respectively, with expected
outputs produced by reference implementations written in Python and C
--- together with nine long-running tasks per suite used for time measurements.

\subsection{Cost and specification/code/proof size}\label{eval-size}

Implementing and proving the three programs cost roughly \$360 with
Claude Fable 5.1, distributed very unevenly (Table~\ref{tab:cost}):
LZ77 alone accounts for two thirds of it, most of it being its optimality
argument. A further \$666 went on optimization, most of which was for the
virtual machine.

\begin{table}
  \caption{\textbf{Content lines (non-blank, non-comment) per artifact.} The
    \emph{spec} column is the Djinnlang specification (the only text a
    human writes/reviews) and is identical for both implementers.}
  \label{tab:size}
  \small
  \begin{tabular}{lrlrrrr}
    \toprule
    & & & \multicolumn{2}{c}{Claude Fable 5.1} & \multicolumn{2}{c}{GPT~6 Astra} \\
    \cmidrule(lr){4-5}\cmidrule(lr){6-7}
    Program & spec & Stage & code & proof & code & proof \\
    \midrule
    Sudoku  &  \textbf{70} & baseline  &  303 &  929 &  76 & 348 \\
            &              & optimized &  448 & 1274 & 197 & 468 \\
    LZ77    &  \textbf{81} & baseline  &  448 & 2339 &  72 & 206 \\
            &              & optimized &  545 & 2585 & 210 & 356 \\
    Imp VM  & \textbf{166} & baseline  &  257 &  655 & 187 & 395 \\
            &              & optimized & 1440 & 2564 & 359 & 514 \\
    \bottomrule
  \end{tabular}
\end{table}

\begin{table}
  \caption{\textbf{Performance of the solutions.} Fastest of repeated runs, in seconds, on identical inputs;
    three runs per cell for workloads under two minutes, two above.
    ${>}900$ is a timeout at the cap; $\dagger$ is a crash (stack
    overflow) after about 0.1~s. All builds are Dafny release builds, tested on a MacBook Pro, M4 Max, 128 GB memory, 16 cores, macOS 15.7.7. Dafny 4.11.0, .NET SDK 9.0.109.}
  \label{tab:perf}
  \small
  \begin{tabular}{llrrrr}
    \toprule
    & & \multicolumn{2}{c}{Claude Fable 5.1} & \multicolumn{2}{c}{GPT~6 Astra} \\
    \cmidrule(lr){3-4}\cmidrule(lr){5-6}
    Program & Workload & base & opt & base & opt \\
    \midrule
    Sudoku & hard puzzles      & 0.65 & 0.13 & ${>}900$ & 47.84 \\
           & 3000 unique       & 0.40 & 0.31 & ${>}900$ &  4.31 \\
    LZ77   & 4\,MB text        & 0.91 & 0.63 & $\dagger$ & 25.63 \\
           & 2\,MB DNA         & 0.33 & 0.24 & $\dagger$ & 28.55 \\
           & 1\,MB periodic    & 0.32 & 0.12 & $\dagger$ &  3.26 \\
    Imp VM & $10^9$ steps      & 76.9 & 18.5 & 211.7 & 189.3 \\
           & Ackermann(3,10)   & 27.7 &  6.5 & 787.6 & 714.5 \\
           & sieve to $2\cdot10^7$ & 23.6 & 6.6 & ${>}900$ & 61.7 \\
    \bottomrule
  \end{tabular}
\end{table}

\begin{table*}
  \caption{\textbf{Wall-clock time and cost.} \emph{impl} is
    implementation and proof to a first accepted solution, \emph{opt}
    the single optimization round. Wall clock is elapsed time for that
    stage; the three programs were compiled concurrently by both
    implementers. Output tokens include
    reasoning. Costs are list prices for the respective model and are
    not directly comparable: our figure is dominated by cache writes
    at the one-hour retention we used, a category the Codex logs do
    not record. For both, we used the subscriptions, so we have not actually spent
    this cost. It is the theoretical API cost.
    Every row covers implementation and optimization only.}
  \label{tab:cost}
  \small
  \begin{tabular}{llrrrrcrrrr}
    \toprule
    & & \multicolumn{4}{c}{Claude Fable 5.1 (Claude Code)} & & \multicolumn{4}{c}{GPT~6 Astra (Codex)} \\
    \cmidrule(lr){3-6}\cmidrule(lr){8-11}
    Program & Stage & turns & output & wall & cost & & turns & output & wall & cost \\
    \midrule
    Sudoku & impl &  315 & 213\,k &  1.8\,h & \$51  & &  51 & 15.6\,k & 13\,m & \$4.31 \\
           & opt  &  154 & 177\,k &  1.7\,h & \$36  & &  58 & 13.1\,k &  7\,m & \$5.74 \\
    LZ77   & impl &  794 & 625\,k &  6.3\,h & \$236 & &  80 & 31.3\,k & 23\,m & \$8.87 \\
           & opt  &  313 & 300\,k &  1.9\,h & \$84  & &  71 & 19.8\,k &  8\,m & \$10.19 \\
    Imp VM & impl &  304 & 116\,k &  1.3\,h & \$73  & &  28 & 12.6\,k & 13\,m & \$3.02 \\
           & opt  &  943 & 763\,k & 12.1\,h & \$546 & &  61 & 19.8\,k & 13\,m & \$7.37 \\
    \midrule
    \multicolumn{2}{l}{all three}
                  & 2823 & 2.2\,M &  & \$1026 & & 416 & 127\,k &  & \$46.89 \\
    \bottomrule
  \end{tabular}
\end{table*}

The two implementers show very different characteristics. Codex compiled all three programs
in 33~minutes of elapsed time for an estimated \$47, against
21~hours and \$1026 for Claude Code --- it is roughly thirty times faster and twenty times
cheaper.

We find that, to solve the same problem, Codex wrote far less proof, and wrote it
in one pass. What that buys is visible in Section~\ref{eval-perf}. The cheap
compilation produced verified programs that are three to four hundred
times slower, and two of its unoptimized programs cannot complete the
workloads within reasonable timeouts. 

Compared against a conventional compiler the numbers are
absurd: no compiler takes six hours or costs \$236 to translate 81
lines.
However, against conventional verification with human-written proofs,
the same numbers look small. This level of insight about our code at
this speed and cost would have seemed impossible to attain even a few years
ago --- and one might reasonably hope that continued development of LLMs
and harnesses will make things even cheaper and faster very quickly.

\subsection{Performance}\label{eval-perf}

As we have explained above, the unambiguity constraint
only applies to inputs and how they relate to outputs. We do not
specify performance, so it is up to the LLM, and we imagine an optimizer
mode that hillclimbs on performance.

Table~\ref{tab:perf} reports the speed of our three example programs.
Again, we find that the two implementers' verified programs differ by a lot.
On the hardest Sudoku workload, Claude's optimized solver takes
0.13~s and Codex's takes 47.8~s; on 4~MB of text our encoder takes
0.63~s and theirs 25.6~s. The virtual machine is closer but still
decisive: 18.5~s against 189~s on a billion interpreter steps, which
is 53~million versus 5~million steps per second.

Each of the implementers was then given one optimization round. Claude Code moved the
virtual machine from 5 to 13 to 53~million steps per second across the
naive, fast, and optimized machines. Codex's round improved
heap-heavy programs substantially --- matrix multiplication from 217
to 38~seconds, the sieve from a timeout to 62 --- but barey changed the others.

Much of this is an artifact of how ambitious the implementers were: Claude Code
seemed to optimize much more for generated code speed, while Codex was faster,
but produced slower code. It is unclear whether Codex would have caught up given
the same time budget. This insight is still useful: Codex could be a
\emph{debug mode} of sorts with higher compile speeds.

\section{Related Work}\label{related-work}

\paragraph{Specification languages.}
Djinnlang belongs to a long tradition of model-based specification
languages. Z~\cite{spivey1992z}, VDM~\cite{jones1990vdm}, and
B~\cite{abrial1996bbook} pioneered the specification of software as mathematical
predicates over abstract states; TLA+~\cite{lamport2002specifying}
specifies concurrent systems as temporal-logic formulas; and
Alloy~\cite{jackson2012alloy} made lightweight relational
specifications automatically analyzable. In all of these, a
specification describes a system but is not, in general, executable:
obtaining an implementation requires manual coding, or manual
machine-checked refinement as in B. Unattended implementation of
a spec is only enabled by the unambiguity constraint and tight integration
with an LLM that we propose.
More recent work proposes tooling that helps humans develop better specifications.
Miyazono et al. set natural-language documentation beside
its Lean formalization with a line-by-line mapping, colour-coding the
passages that correspond or diverge, using Signal's X3DH as their case
study~\cite{miyazono2025specide}. Dodds argues
that most systems have no coherent formal specification at
all, only partial and mutually contradictory ones, and that writing
specifications is itself a programming-like activity in want of
tools~\cite{dodds2025specs}.

\paragraph{Contracts and verification-aware languages.}
Eiffel made pre- and
postconditions part of the program text and of the methodology
(``design by contract'')~\cite{meyer1992dbc}, followed by
JML for Java~\cite{leavens2006jml} and
Spec\#~\cite{barnett2004specsharp}. Verification-aware languages ---
Dafny~\cite{leino2010dafny}, F*~\cite{swamy2016fstar}, and
Verus~\cite{lattuada2023verus} --- statically verify the code against
its contracts. In all of these the programmer writes both the
specification and the implementation, and nothing forces the
specification to be complete. Djinnlang keeps the contract vocabulary
(indeed it compiles to Dafny and reuses its types and verifier) but
removes the human-written implementation entirely, and its
unambiguousness proofs remove the option of a partial spec.

\paragraph{Synthesis from specifications.}
Deriving implementations from specifications symbolically is a classic
ambition and could thus be considered an instance of
\emph{spec-driven
development}~\citep{martinelli2026}.
The refinement calculus~\cite{morgan1994programming} treats
specifications and code as points in one ordering, refined by hand;
Fiat~\cite{delaware2015fiat} mechanizes such derivations in Rocq;
complete functional synthesis~\cite{kuncak2010complete} compiles
constraints over decidable theories directly to code; and
Sketch~\cite{solarlezama2008sketching} fills holes in partial programs
by combinatorial search. These systems are predictable and sound but
restricted --- to decidable theories, to finite search spaces, or to
domains with hand-built refinement libraries. The Verse
calculus~\cite{augustsson2023verse} meets the same limit from the
language side, leaving the evaluation strategy that would solve its
equations to future work (see Section~\ref{concepts}).
Closest to our aims is relational
compilation~\cite{pitclaudel2022rupicola}, which recasts extraction as
proof search and derives fast low-level code from
annotated functional models, emitting a proof for every compilation.
They automate only the descent from a functional model to code,
leaving that model's agreement with an abstract specification to the
programmer. Djinnlang uses a similar core architecture, but uses LLMs for
synthesis. We thus trade symbolic search for an LLM's generality while keeping the soundness:
a candidate implementation that does not provably meet the spec is
rejected. 

\paragraph{LLMs for verified code and proofs.}
A fast-growing body of work polices LLM output with verifiers.
VerMCTS guides tree search over Dafny and Rocq programs with a verifier
in the loop~\cite{brandfonbrener2024vermcts};
Clover checks code--spec--docstring consistency~\cite{sun2024clover};
AlphaVerus bootstraps verified Verus code without human
labels~\cite{aggarwal2025alphaverus}; and
AutoVerus~\cite{yang2025autoverus} and Baldur~\cite{first2023baldur}
generate proofs for existing code, as do LLM-based invariant
inference~\cite{kamath2023invariants} and specification
generation~\cite{ma2025specgen,misu2024towards}.
DafnyBench measures LLMs' ability to complete Dafny
verifications at scale~\cite{loughridge2024dafnybench}, and Formal
Disco generates open-ended synthetic verified programs in Dafny,
Verus, and Frama-C to train models for verification
tasks~\cite{poesia2026formaldisco}. This
literature treats the LLM as an assistant working on or alongside
conventional programs. Djinnlang inverts the workflow: the human never
writes code, the LLM-plus-verifier pair is packaged as a compiler, and
--- uniquely, to our knowledge --- the unambiguity constraint
guarantees the LLM has no semantic leeway, so its output need never be
read.

\paragraph{Specification elicitation problem.}
A recent critical strand argues that neither proofs nor specifications
survive contact with LLMs unexamined. Von Hippel et
al. argue that the bottleneck in secure program synthesis is the
specification itself, a human-computer interaction problem that
stronger models do not solve~\cite{vonhippel2026sps, dougherty2026lies}. Lahiri
quantifies it, symbolically testing whether a specification captures
user intent at all~\cite{lahiri2024intent}. Djinnlang does not fully resolve
this problem either; but narrows the trusted artifact to the specification, and
aims to make it easier to read and less error-prone (see Section~\ref{eval-redundancy}).

\section{Limitations}\label{limitations}
The limitations of the approach fall into three groups: what a
compilation costs, what the language cannot express, and how far the
guarantee reaches.

\paragraph{Compilation cost and latency.}
Compilation expenses are measured in hours and dollars rather than seconds.
Implementing and proving the three programs of
Section~\ref{eval-empirical} cost roughly \$360 in model usage with
Claude Fable~5.1, the longest single program taking 6.3~hours; a
further \$666 went on optimization. Compiling all three concurrently
took 20.8~hours of elapsed time and \$1026 in total
(Table~\ref{tab:cost}).
Djinnlang especially suits artifacts written once and depended on for a long
time.

\paragraph{Non-convergence.}
The worst outcome of a Djinnlang compilation is not a rejected
implementation but no answer. When
the model neither finds a proof nor a
counterexample, the programmer is handed only a timeout.
The three explanations --- the
specification is unsatisfiable, the proof is merely hard, the model is
too weak --- are indistinguishable from outside.
We did not in fact encounter this outcome in any of the four programs
we compiled.

\paragraph{Concurrency.}
Djinnlang is not thread-safe, has no shared mutable state, and no interleaving.
Specifying concurrency needs temporal properties, as implemented in
TLA+~\cite{lamport2002specifying}, and reconciling those with
unambiguity is an open problem.

\paragraph{Side-channel security.}
An implementation
can be fully verified and still leak a key through timing
(Section~\ref{unambiguity}). Constant-time execution is a performance detail
that Djinnlang hides by design.

\paragraph{Soundness bugs in the verification pipeline.}
A developer must trust the ambient Dafny
supplying the datatypes and helpers a spec is written against
plus the external I/O bindings, the
unverified \texttt{main} glue, the pinned drivers, the translator, as well as
Dafny, Boogie, and Z3.

\section{Future Work}\label{future-work}
An abstract goal of this work is to allow a programmer to express their
intent in a shorter, elegant, and less error-prone way, and we hope
that our unambiguity constraint and LLM-in-the-compiler paradigm will allow for such
advances in the future. Concretely, we suggest the following areas of further study:

  \paragraph{A fine-grained effect system.} Effects are currently a thin
  layer meant for interoperability with the imperative world: a \texttt{proc}
  says \emph{which} effects happen and in what order, but the language
  cannot say much \emph{about} them. Effects could be much more. We
  want constraints over the event trace, written in
  the same constraint language as everything else: for instance ``every file this
  proc opens is eventually closed,'', ``we print a grid and thus never lines longer than 40 characters'', ``no network traffic occurs after
  the credential is read,'' We expect this will make systems-style
  code much more elegant.

  \paragraph{Principled relaxations of unambiguity.} Unambiguity should
  remain the default, but we might want some leeway: take the
  exact wording of an error message, or the choice among 
  different but equivalent library calls. As a first step, this could
  be implemented on top of procedures, where it would be easy to call
  into some function the LLM has freely generated.
  Relaxing unambiguity on relational parts of the language could follow.

  \paragraph{Interactive tooling.}
  There should be good, pleasant-to-use tooling for writing specifications.
    For instance this tooling could help a developer gain insights about a spec before
    the LLM implements it. 
A bounded checker over the
  lowered relation, in the spirit of Alloy's
  analyzer~\cite{jackson2012alloy}, could try to find counterexamples while
  the user is typing (Section~\ref{limitations}).
    The front end is symbolic and fast, so a language server can report lexing,
  layout, parsing, check errors, and provide progress details on a background process
  that writes the proofs. Proofs and implementations can be cached
  and shared across runs. 

  \paragraph{Specifying performance and scaling characteristics.} The LLM implementer currently owns
  all performance decisions; the verifier checks only functional
  correctness. As noted, this is desirable ---
  LLMs are good optimizers --- but unacceptable for timing-sensitive
  code and a liability for DoS-style worst cases. We plan simple syntax
  for asymptotic complexity bounds and baseline-performance
  constraints, checked either by verified resource analysis or, more
  pragmatically, by compiler-generated benchmark harnesses.

\section{Conclusion}\label{conclusion}

We presented the unambiguity constraint and LLM-in-the-compiler paradigm, and showed
their partical implementation on Djinnlang, a high-level specification language in which the
programmer writes only specifications and an LLM handles implementation and proof. 

Because every function's spec must be proven unambiguous,
the LLM has no leeway on program semantics, the generated code never
needs to be read, and it can be regenerated from the spec at any time.
Under this constraint, the distinction between a specification
language and a programming language essentially disappears.

We have shown that specs can be shorter and closer to intent than the code they replace,
and help catch errors. Finally, we have shown that LLMs are able to do the required
proofs and implementation, albeit currently at high cost.

More broadly, this work is evidence for an underexplored role for LLMs
in language design. Used not as an autocomplete or unattended vibe-coder
but as a compiler stage and tightly constrained by a verifier,
LLMs make abstractions viable that no symbolic compiler could
support: programming by unambiguous specification. This constraint both
makes this a compiler and acts as an AI control mechanism. It can help
us reach a world of programming in which humans say \emph{what} and machines decide
\emph{how}.

\newpage

\section*{Acknowledgments}
We thank Will Byrd, Jake Ginesin, Gabriel Poesia, Max von Hippel, Paul Krogmeier, Cameron Wong, Raffi Sanna, Rhea Karty, Yilun Du, Mahzarin Banaji, Samuel Le{\ss}mann, and Trevor DePodesta for their feedback.
This research was partially supported by a Coefficient Giving grant, funding and credits from
an Amazon Research Award, and a Harvard Data Science Initiative grant. Simon Henniger was supported by
the Harvard John A. Paulson School of Engineering and Applied Sciences (SEAS) Prize Fellowship
and the German Academic Fellowship Organization, funded by the German Federal Ministry for
Economic Affairs and Energy.

\bibliographystyle{plainnat}
\bibliography{paper}

\end{document}